# Do we need the g-index?

**Michael Schreiber**

*Institute of Physics, Chemnitz University of Technology, 09107 Chemnitz, Germany. E-mail: schreiber@physik.tu-chemnitz.de*

Using a very small sample of 8 datasets it was recently shown by De Visscher (2011) that the g-index is very close to the square root of the total number of citations. It was argued that there is no bibliometrically meaningful difference. Using another somewhat larger empirical sample of 26 datasets I show that the difference may be larger and I argue in favor of the g-index.

1.  **Introduction**

The h-index is given by the largest number $h$ of papers of a researcher which have received at least $h$ citations. One criticism is that the excess citations, i.e., all citations which exceed $h$ for a given publication do not have an effect. Therefore Egghe (2006) proposed the g-index which is given by the largest number $g$ of papers which have received at least $g$ citations on average (Schreiber, 2010). Recently De Visscher (2011) utilized a rather small sample of the citation records of 8 famous physicists which had previously been analyzed by Schreiber (2008a, 2010). De Visscher showed that $g$ is very close to the square root $\sqrt{S}$ of the total number $S$ of citations which would mean that the index is essentially a metric of this total number. De Visscher concluded that the g-index is not necessary and he also argued that the h-index is more meaningful, because it rewards a so-called "consistent" citation record in which the highly cited papers show a rather homogeneous citation distribution. Egghe (2012) took a different view and argued in favor of the g-index, just because it awards non-consistent impact.

De Visscher already pointed out that it is premature to generalize his results because of the small sample size and he proposed further studies. It is the aim of the present investigation to provide such an analysis for a different and larger sample, comprising the citation datasets of 26 more average physicists. In this case the difference between $g$ and $\sqrt{S}$ is more pronounced as already suspected by De Visscher (2011). This can be traced back to the fact that a larger share of the papers does not contribute to the g-core, i.e., the g-defining set of publications. Although these papers have rather small citation frequencies, their cumulative effect is not so small that the deviations between $g$ and $\sqrt{S}$ can be neglected. Additionally the observation that certain datasets yield especially large deviations corroborates my assumption that there is a bibliometrically meaningful difference between $g$ and $\sqrt{S}$. These outliers occur when the papers in the long tail of the citation distribution beyond the value of the g-index comprise a rather large number of citations.

2.  **Data and definitions**

The citation records of the 8 highly cited physicists have been obtained from the Web of Science in July 2007 (Schreiber, 2008a) and used for an investigation of the g-index (Schreiber, 2010). For the 26 physicists from my home Institute of Physics at Chemnitz University of Technology I have harvested the Web of Science in January and February 2007 (Schreiber, 2007) and analyzed these data with respect to the g-index, too (Schreiber, 2008b). All datasets have been carefully screened with respect to homonyms in order to establish the integrity of the raw data.

The citation record of each scientist is sorted in decreasing order in terms of the number $c(r)$ of citations for each paper, so that $r$ is the rank of the paper. The h-index is then defined as the largest rank $h$ for which

$$c(h) \geq h . \qquad (1)$$

If one denotes the average number of citations as

$$\bar{c}(r) = \frac{1}{r} \sum_{i}^{r} c(i), \qquad (2)$$

then the g-index can be defined as the largest rank $g$ for which

$$\bar{c}(g) \geq g. \qquad (3)$$

This is equivalent to the original definition by Egghe (2006) in terms of the sum of citations

$$s(r) = \sum_{i=1}^{r} c(i) = r\,\bar{c}(r), \qquad (4)$$

namely that $g$ is the largest rank for which

$$s(g) \geq g^2. \qquad (5)$$

If $n$ is the total number of papers, we get $S = s(n)$ as the total number of citations.

Using a straightforward piecewise linear interpolation of the rank-frequency function $c(x)$ (Schreiber, 2007, 2008b; Guns & Rousseau, 2009) one can avoid the inequalities and define the interpolated h-index $\tilde{h}$ by

$$c(\tilde{h}) = \tilde{h} \qquad (6)$$

and the interpolated g-index $\tilde{g}$ by

$$\bar{c}(\tilde{g}) = \tilde{g} \qquad (7)$$

or, equivalently, by

$$s(\tilde{g}) = \tilde{g}^2. \qquad (8)$$

Similarly, the A-index gives the average number of citations in the h-core

$$\bar{c}(h) = A \qquad (9)$$

and

$$\bar{c}(\tilde{h}) = \tilde{A} \qquad (10)$$

and the R-index denotes the square root of the number of citations in the h-core

$$s(h) = R^2 \qquad (11)$$

and

$$s(\tilde{h}) = \tilde{R}^2. \qquad (12)$$

The advantage of the interpolated indices is that not so many ties occur in the evaluation, i.e., there are not so many coinciding values of the indices for different datasets. Moreover, inconsistencies due to the discreteness which can occur for small datasets are avoided, for example we always have $R \leq \tilde{R} \leq \tilde{g}$ in contrast to the example given by Egghe (2012) in which $R > g$ in contradiction to the formal result.

The actual data of the indices have been given elsewhere for the 8 datasets of highly cited physicists (Schreiber, 2008a) as well for the 26 datasets of more average physicists (Schreiber, 2007, 2008b) and shall not be repeated here.

### 3. Results and discussion

In Figure 1 the obtained values for $\tilde{g}$ are displayed in dependence on the square root of the total number of citations, $\sqrt{S}$, for the 26 average physicists. The data show some scatter around the line which was obtained by simple linear regression analysis ($R^2 = 0.951$). It is obvious from this display that a ranking of the scientists according to $\tilde{g}$ yields different results compared to a ranking in terms of $\sqrt{S}$. In my view this already means that there is a meaningful difference between these two measures. Moreover, I note that the steepness of the regression function is significantly smaller than the gradient of the respective line for the 8 highly cited physicists which is shown in Figure 2, where the data are displayed on a doubly logarithmic scale for easier distinction. The data points for those 8 researchers lie significantly beyond the range displayed in Figure 1, and the linear regression for those 8 data yields a line which is above most of the 26 data points in Figure 1.

It is worthwhile to note that these regression lines have been determined so that they have no offset, i.e., they run through the origin $\tilde{g} = \sqrt{S} = 0$. In my view this makes more sense than the regression computed by De Visscher where an offset was allowed. I do not think that such an offset is meaningful, since negative values of $g$ are not possible and already for $S = 1$ we have $g = \tilde{g} = 1$. However, I find it interesting to note that the offset of -3.04 as determined by De Visscher from the sample with 8 datasets yields a reasonably good representation of the respective 26 data points in Figure 1.

De Visscher (2011) found that for the sample of the 8 datasets of highly cited physicists most values of the g-index were very close to $\sqrt{S}$. The two observed outliers with $g \ll \sqrt{S}$ showed a high productivity coupled with a not so strongly skewed citation distribution. It is a matter of discussion whether the resulting smaller values of the g-index should be preferred (Egghe, 2012) or not (De Visscher, 2011). But it is clearly a meaningful difference between the g-index and $\sqrt{S}$ which should not be easily discarded. For the sample of 26 datasets there are so many "outliers" above and below the regression line that it is obvious that the g-index is not simply a metric of $\sqrt{S}$.

In Figure 3 the data for the index $\tilde{R}$ are presented. As observed previously (Schreiber, 2008b), they do not differ very much from the values for $\tilde{g}$, because the definition of $\tilde{R}$ in (12) is similar to that of $\tilde{g}$ in (8), but only based on a smaller number of papers. Thus $\tilde{R}$ awards a strongly skewed citation distribution even more than the g-index. The calculation of the regression function yielded a correlation with $\sqrt{S}$ of $R^2 = 0.943$.

Noting the similarity between the definitions of $\tilde{g}$ in (7) and $\tilde{A}$ in (10), one might as well compare $\tilde{g}$ and $\tilde{A}$ in Figure 3. Again $\tilde{A}$ is based on a smaller number of papers and thus favours authors with a strongly skewed citation distribution. But the correlation with $\sqrt{S}$ is not so strong, $R^2 = 0.832$.

The data for the index $\tilde{h}$ in dependence on $\sqrt{S}$ are also presented in Figure 1 as well as the corresponding linear regression ($R^2 = 0.962$). As observed by De Visscher the correlation is rather large. And there seems to be a trend of the lower values of the index to be above the regression line and the larger values to fall below the line. Of course, this would change, if an offset were allowed in the linear regression. Again the regression function based on the data for the 8 highly cited physicists is significantly different, but this time it is less steep, see Figure 2. Thus 24 of the 26 data points are above the respective line. It is worthwhile to note that the regression with an offset of 9.8 as determined by De Visscher for the 8 datasets does not result in a reasonable representation of the 26 data points in Figure 1. In any case, the offset is not appropriate, because $h$ can never be larger than $\sqrt{S}$. Therefore the wedge for small values of $\sqrt{S}$ in Figure 1 between the diagonal and the regression function determined by De Visscher for the h-index does not make sense. Moreover, we always have $\tilde{g} > \tilde{h}$ which means that the even larger wedge between the two respective regression functions is unreasonable.

## 4. Concluding remarks

The analysis of De Visscher (2011) was based on a very small sample of the citation records of 8 highly cited physicists. The present investigation of 26 citation records of more average physicists yielded similar results. But the deviations between the g-index and the square root of the total number of citations turned out to be somewhat larger. The reason is that the long tail of lowly cited publications has a bigger influence on $\sqrt{S}$. In conclusion it may not be a good idea to discard the g-index in favor of $\sqrt{S}$. In my view those lowly cited publications should not be taken into account. There is another practical reason for this point of view: It is much more difficult to establish the integrity of the citation record, if all lowly cited publications have to be taken into consideration. One has to check many more papers.

It is not surprising that differences have been found between the g-index and $\sqrt{S}$. These differences should not be taken as a justification of the need for both indicators, because then one might be tempted to argue for an ever increasing number of indices and other indicators. With regard to the aim of measuring the scientific impact of the publications of a researcher, most people probably agree that such indices are quite noisy indicators with a substantial degree of intrinsic uncertainty concerning the scientific impact. Although the numbers of the various indices can be calculated with good accuracy, this precision is misleading, because relatively small differences between the values of the same index for different researchers should never be utilized to value one researcher better or worse than the other. In other words, the interpretation of a bibliometric index should always be done with great care. As a consequence, it is not necessary to make use of two (or more) indices that correlate quite strongly and therefore I argue that the indicator $\sqrt{S}$ can be discarded in favour of the g-index. Nevertheless, $\sqrt{S}$ has some usefulness in detecting the outliers discussed in the previous section, i.e. researchers characterized by a high productivity with the consequence that citations are scattered over a larger number of papers rather than concentrated on a few core publications of these researchers. In my opinion fewer papers with more citations are preferable and this point of view certainly influences my preference of the g-index in favour of $\sqrt{S}$.

In comparison with the h-index I agree with the divergent vision expressed by Egghe (2012), namely that it is preferable to award a higher index to a non-homogeneous citation record as it is done by the g-index in comparison with the h-index. However, from a practical point of view the significantly larger values of the g-index in comparison with the h-index also mean that a significantly larger effort is necessary to establish the data base. This might be overcome by generalizing the g-index (van Eck & Waltman, 2008) which is easily done by introducing a prefactor on the right hand sides of Equations (3), (5), (7), and (8). In this way the arbitrariness of the definition of the g-index is exploited in the same way as it can be done for the h-index (van Eck & Waltman, 2008; Schreiber, 2013). For example a prefactor of 2 would yield significantly smaller values which would be of the same order of magnitude as the values of the h-index. Thus the effort in establishing the data base would be also of the same order of magnitude and therefore the practical disadvantage of the g-index would be avoided.

In summary, I conclude that we may not need the g-index just as we do not really need the h-index as a one-dimensional measure of scientific impact. However, in my view the g-index should not be discarded to the benefit of $\sqrt{S}$, but it should rather be favored in comparison with $\sqrt{S}$ as well as with $h$. And the answer to De Visscher's question "What does the g-index really measure?" therefore is: It certainly measures the impact of the productive core, and the outliers in Figure 1 show meaningful deviations from the total impact as represented by (the square root of) the total number of citations.

**References**

De Visscher, A. (2011). What does the g-index really measure? *Journal of the American Society for Information Science and Technology* 62(11), 2290-2293.

Egghe, L. (2006). Theory and practice of the g-index. *Scientometrics* 69(1), 131-152.

Egghe, L. (2012). Remarks on the paper of A. De Visscher "What does the g-index really measure?" *Journal of the American Society for Information Science and Technology* 63(10), 2118-2121.

Guns, R. & Rousseau, R. (2009). Real and rational variants of the h-index and the g-index. *J. Informetrics* 3, 64-71.

Schreiber, M. (2007). A case study of the Hirsch index for 26 non-prominent physicists. A*nnalen der Physik (Leipzig)* 16(9), 640-652.

Schreiber, M. (2008a). To share the fame in a fair way, $h_m$ modifies $h$ for multi-authored manuscripts. *New Journal of Physics* 10, 040201–1-9.

Schreiber, M. (2008b). An empirical investigation of the *g*-index for 26 physicists in comparison with the *h*-index, the *A*-index, and the *R*-index. *Journal of the American Society for Information Science and Technology* 59(9), 1513-1522.

Schreiber, M. (2010). Revisiting the g-index: the average number of citations in the g-core. *Journal of the American Society for Information Science and Technology* 61(1), 169-174.

Schreiber, M. (2013). A case study of the arbitrariness of the h-index and the highly-cited-publications indicator. *Journal of Informetrics* (accepted).

Van Eck, N.J. & Waltman, L. (2008). Generalizing the h- and g-indices. *Journal of Informetrics* 2(4), 263-271.

**Fig. 1.** Values of the $\tilde{h}$-index and the $\tilde{g}$-index in dependence on the square root of the total number of citations, $\sqrt{S}$, for the sample of 26 average physicists; the thin solid line is the diagonal, thick solid lines show the respective regression functions (without offset), dotted lines are the regression functions with offset determined by De Visscher (2011) for the sample of 8 highly cited physicists.

**Fig. 2.** Same as Figure 1, but on a doubly logarithmic scale, including the data points of the 8 highly cited physicists which have been analyzed by De Visscher (2011) and the respective regression functions (without offset) shown by broken lines. Note that the linear regression functions with offset do not yield straight lines in the doubly logarithmic plot. The regression functions are determined for the original data, not for the logarithmized values.

**Fig. 3.** Values of the $\tilde{h}$-index, the $\tilde{R}$-index, the $\tilde{A}$-index, and the $\tilde{g}$-index in dependence on the square root of the total number of citations, $\sqrt{S}$, for the sample of 26 average physicists; solid lines show the respective regression functions (without offset).

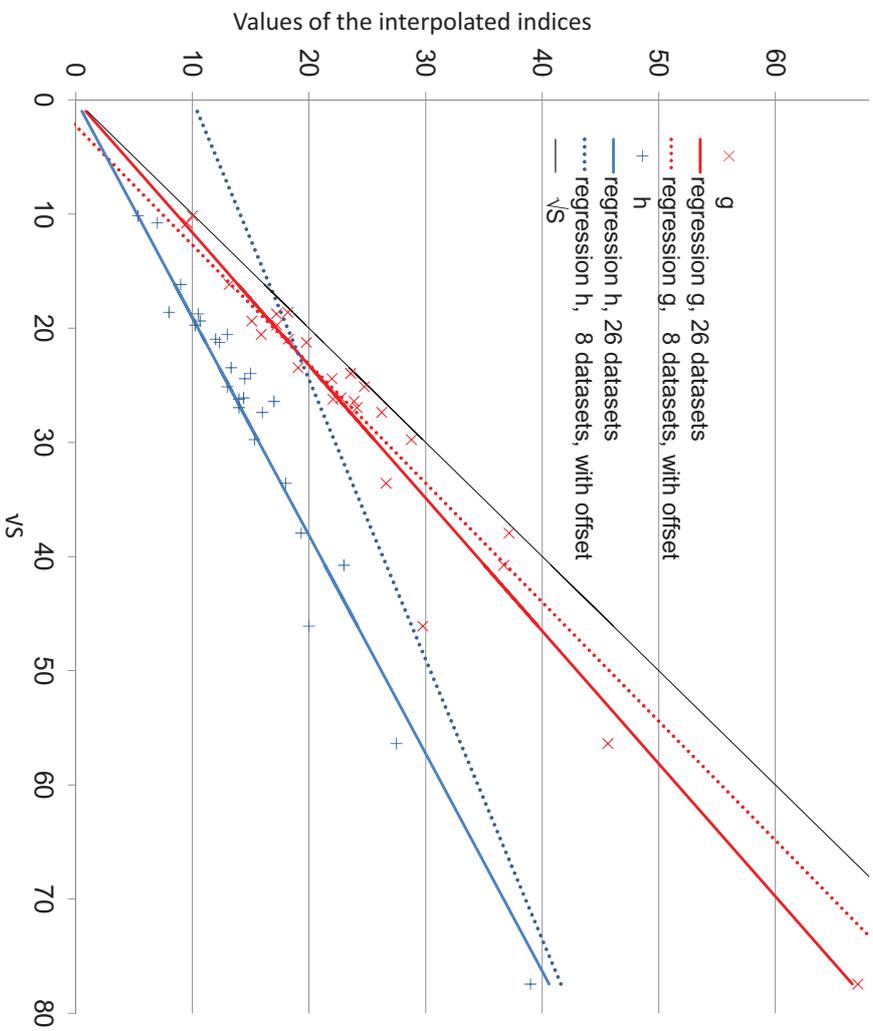
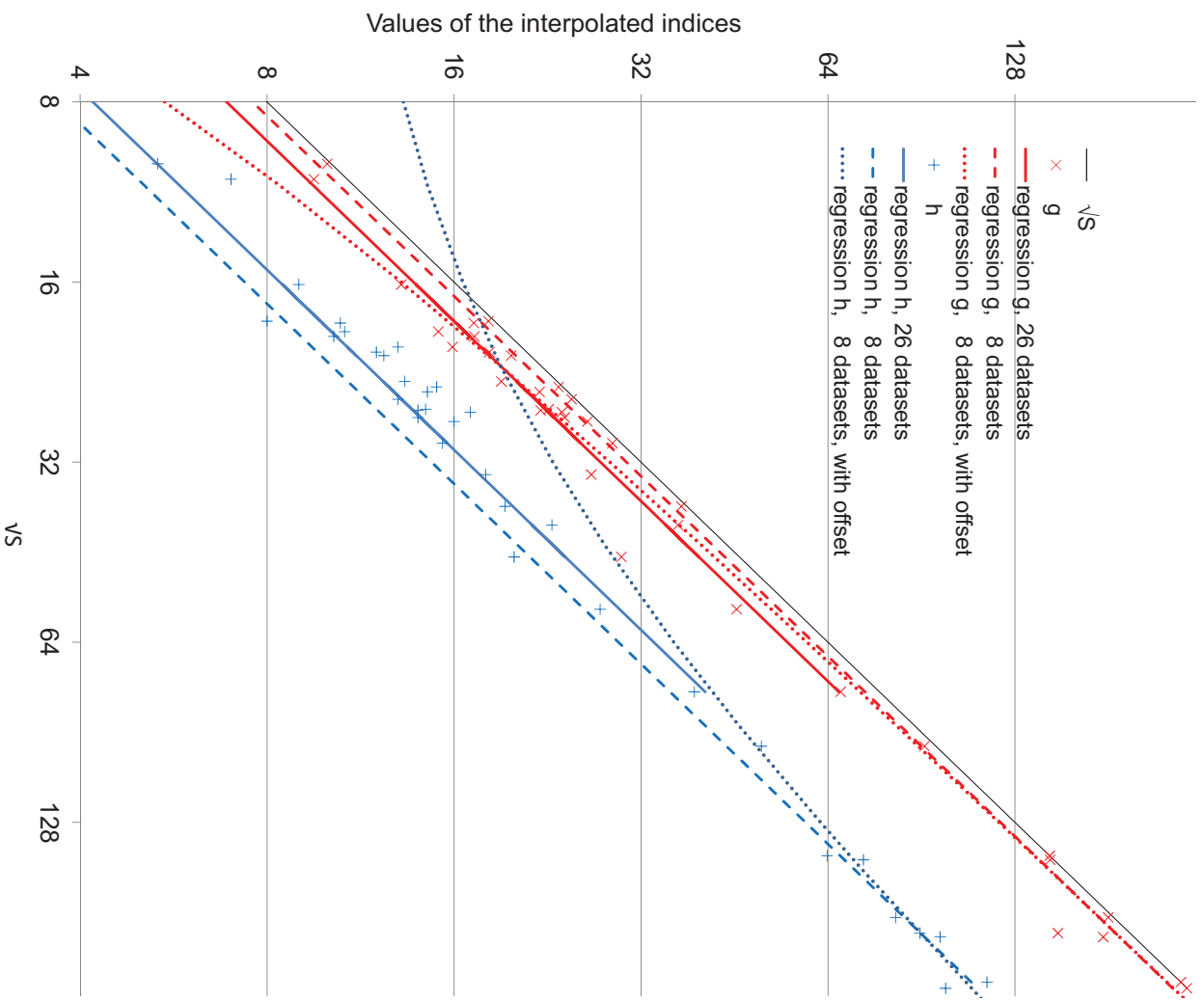

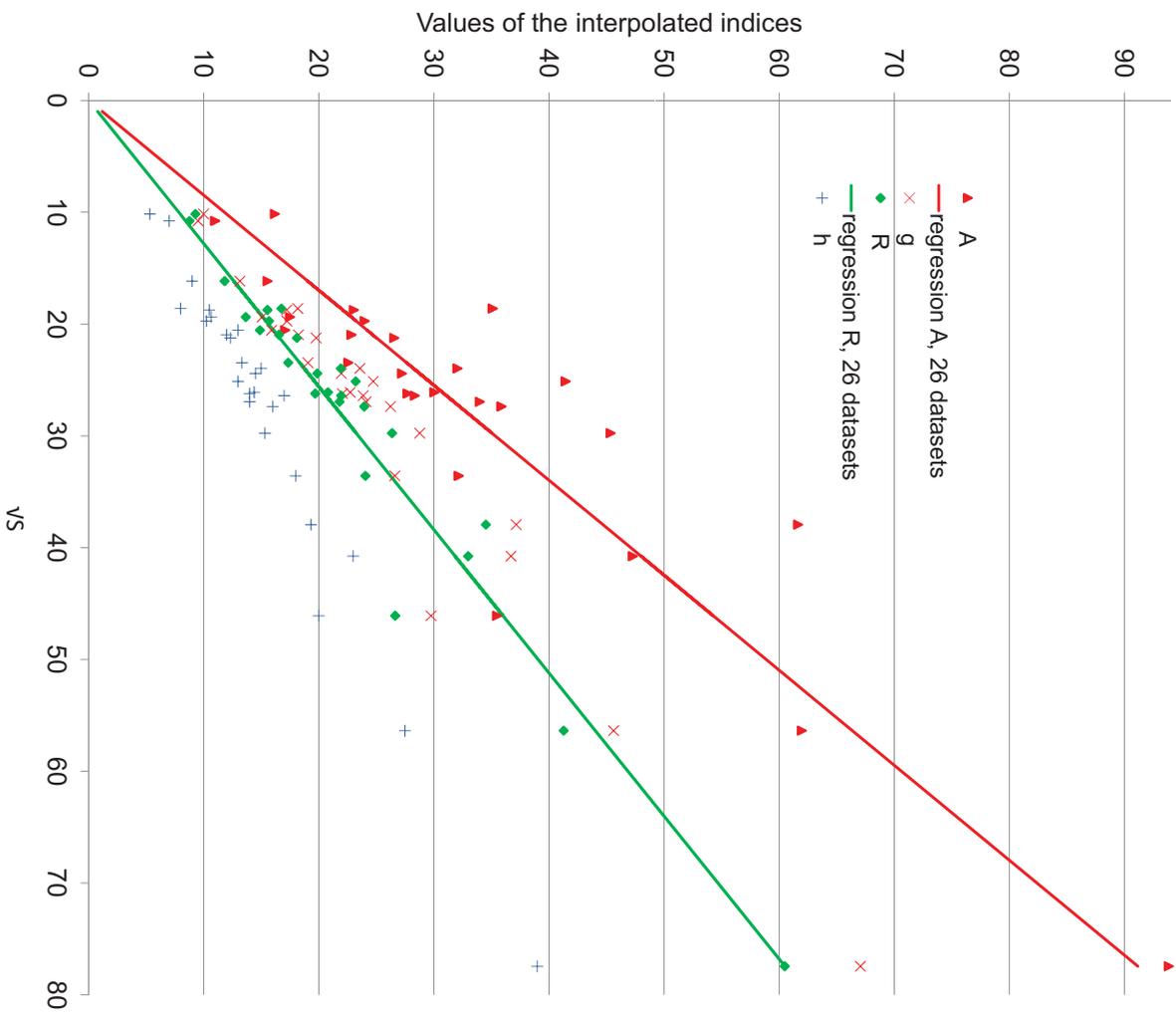